\documentclass[12pt]{article}

\usepackage[margin=1in]{geometry}

\usepackage{graphicx}

\usepackage{amsmath,amssymb}

\usepackage{authblk}

\usepackage{hyperref}

\title{Design of a quantum diamond microscope with efficient scanning confocal readout}

\author[1]{Daniel G. Ang\thanks{These authors contributed equally to this work.}}

\author[1,2]{Jiashen Tang\thanks{These authors contributed equally to this work.}}

\author[1,2,3]{Ronald L. Walsworth\thanks{Correspondence: Daniel Ang (dga@umd.edu), Ronald Walsworth (walsworth@umd.edu).}}

\affil[1]{Quantum Technology Center, University of Maryland, College Park, Maryland 20742, USA}
\affil[2]{Department of Physics, University of Maryland, College Park, Maryland 20742, USA}
\affil[3]{Department of Electrical Engineering and Computer Science, University of Maryland, College Park, Maryland 20742, USA}

\date{}

\begin{document}
\maketitle

% ----------------------------
% Abstract
% ----------------------------
\begin{abstract}
We introduce the light-sheet confocal quantum diamond microscope (LC-QDM) for widefield 3D quantum sensing with efficient confocal readout. The LC-QDM leverages light-sheet illumination and laser scanning confocal methods to enable high-resolution, high-speed 3D measurements with nitrogen-vacancy (NV) defects in diamond, combining the best of widefield and confocal modalities in a single device and eliminating the need for thin-NV-layer diamond chips. We perform simulations and measurements of NV initialization and readout times to model the anticipated performance of the LC-QDM compared to existing QDM designs. Our findings show that the LC-QDM will provide significant advantages for applications requiring limited laser power.
\end{abstract}

% If you want an explicit keywords line
\noindent \textbf{Keywords:} quantum diamond microscope; quantum sensing; nitrogen-vacancy centers

% ----------------------------
% Main Body
% ----------------------------
\section{Introduction}
\label{sec:introduction}

The quantum diamond microscope (QDM) is an established quantum sensing technology that images properties of interest such as magnetic fields, strain, and temperature. The QDM provides a wide field of view (FOV), high spatial resolution, and operation in ambient conditions~\cite{levine_principles_2019,Tang2023QDM,BarrySensOpt2020}. These features have enabled diverse applications in fields such as material science \cite{ku_imaging_2020}, life sciences \cite{barry_optical_2016, fescenko2019diamond,davis2018mapping,Kazi2021wide}, geology \cite{glenn2017micrometer, fu2023pinpointing,mittelholz2024magnetic, taylor2023direct}, microelectronics analysis \cite{turner_magnetic_2020,lenz2024hardware,oliver2021vector,kehayias2022measurement}, and efforts at dark matter detection \cite{marshall_directional_2021,ebadi_directional_2022,marshall_high-precision_2022}. The QDM utilizes an ensemble of negatively-charged nitrogen-vacancy (NV) centers in diamond for quantum sensing~\cite{dohertyNitrogenvacancyColourCentre2013}. Illumination with green (typically 532\,nm) laser light is used for NV initialization and readout of the 
sensor information encoded in the NV photoluminescence (PL, $\sim$600$-$800\,nm). QDM spatial resolution is typically determined by the smallest volume of NV diamond that can be optically differentiated during readout (hereinafter referred to as the sensing voxel) \cite{rietwyk2024practical}. This volume represents the smallest 3D unit of NV centers that can be independently measured. Beyond spatial resolution, accurate source reconstruction also depends on minimizing the distance between the sample and NV sensors \cite{broadway_improved_2020,roth1989using}. 

To achieve $\sim$1 $\mu$m lateral resolution in a QDM, a high-magnification imaging system with a dense camera pixel array is typically employed \cite{Tang2023QDM}. This setup resolves NV sensing voxels in a 2D plane near the sample at the optical diffraction limit. However, because a camera-based imaging system lacks optical Z-sectioning, a thin NV layer (on the nano- to micron-scale) is required to prevent QDM image blurring. Fabricating layered NV diamond chips requires advanced techniques, such as ion implantation or chemical vapor deposition (CVD) at the surface of an existing diamond substrate~\cite{levine_principles_2019}, whereas commercially available quantum-grade diamond chips typically contain NV centers throughout the entire $\sim$\,0.5\,mm-thick bulk diamond substrate \cite{edmonds2021characterisation,thorlabs}.  Additionally, a layered diamond chip limits quantum sensing to the active NV layer. This restricts quantum sensing to a single plane, rendering it unsuitable for applications like gradiometry and others that require flexibility in placement of the sensing layer~\cite{ebadi_directional_2022,tetienneProximityInducedArtefactsMagnetic2018,Ang2024}.

An alternative method to achieve micron-scale QDM resolution is point-by-point detection with a scanning confocal system, which enables readout from each diffraction-limited sensing voxel using a pinhole to reject out-of-focus light~\cite{gruberScanningConfocalOptical1997,nicholsVideorateScanningConfocal2011}. However, moving the interrogation point with galvo mirrors or translation stages introduces dead times ($\gg$1 ms) that far exceed the duration of the quantum sensing protocol (typically $<$1 ms, limited by the NV spin polarization time $T_1$). Recent QDM designs by Leibold et al. \cite{leibold2024time} and Cambria et al. \cite{cambria2024scalable} mitigate this problem by using a pair of acousto-optic modulators (AOMs) to rapidly scan the green laser, reducing the dead time to $\sim$100\,ns. Nonetheless, these systems do not have a pinhole for Z-sectioning capability, and thus they cannot be used with commercial bulk diamond chips. Furthermore, Ref.~\cite{leibold2024time} utilizes a recurrent readout scheme where multiple sensing voxels are interrogated after applying a global MW control sequence applied to a wide FOV, resulting in increased readout duty cycle. In many QDM applications, much more time is spent on NV spin initialization than on applying MW pulses, limiting the efficiency enhancement of this approach.

Here, we introduce a new QDM design that addresses these limitations by acquiring widefield images with a combination of optical-sectioning and highly-efficient confocal readout, thereby offering an upgrade over existing setups that are either limited by slow scanning speed, lack of Z-sectioning, or the need for NV-layered diamond chips. Hereafter, we refer to this system as a light-sheet QDM with confocal readout (LC-QDM). 

The LC-QDM distinguishes itself from previous QDMs~\cite{levine_principles_2019,Tang2023QDM,leibold2024time,cambria2024scalable} due to two innovations. First, a sheet of green light is employed to perform global initialization of all NV sensing voxels within a wide FOV. Second, rapid scanning confocal readout is performed by an optical system consisting of a pinhole and two pairs of AOMs. The LC-QDM's enhanced capabilities are applicable to many QDM applications utilizing a confocal setup, such as for 2D materials characterization~\cite{ku_imaging_2020}, strain imaging~\cite{marshall_high-precision_2022}, super-resolution techniques~\cite{maurer_far-field_spinresolft_2010,jaskula_superresolution_spinresolft_2017,arai_fourier_2015}, and ion-induced damage track detection for sensor optimization~\cite{rackeVacancyDiffusionNitrogenvacancy2021,yamamotoIsotopicIdentificationEngineered2014} and fundamental physics applications~\cite{marshall_directional_2021,ebadi_directional_2022}. As shown in Section~\ref{sec:anticipatedperformance} , the LC-QDM's efficient readout protocol offers a particular advantage for applications requiring limited laser power, such as bioimaging~\cite{LeSage2013,barry_optical_2016}, where reduced photoxocity enables longer, more accurate mapping of magnetic fields in living cells.

\section{Design of the LC-QDM}

\begin{figure}[htbp]
    \centering
    \includegraphics[width=\linewidth]{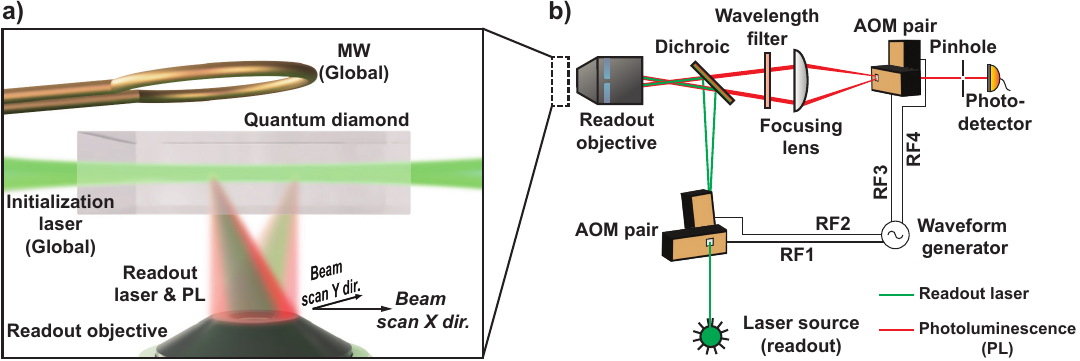}
    \caption{Schematic of LC-QDM with high-speed, diffraction-limited readout. a) Zoom-in view of the NV-diamond sensor head. A green laser beam, shaped into a wide sheet, illuminates a large area of the diamond to initialize NV spins across the entire FOV. A MW pulse sequence manipulates all NVs within the same FOV to perform quantum sensing. Optical readout of the NV sensor is performed point-by-point by focusing a second green laser beam through a microscope objective from below and collecting the resulting red NV PL from each sensing voxel. b) Additional hardware components associated with NV readout. Two pairs of AOMs provide rapid control of the readout laser's propagation direction and the resulting location of induced NV red PL. Synchronizing the RF sources for each AOM pair enables precise XY-pointing of the readout laser and maintains alignment of the red PL with the pinhole, enabling optical sectioning and confocal NV PL readout.
    } 
    \label{fig:1}
\end{figure}

Figure \ref{fig:1} illustrates the LC-QDM's hardware setup, showcasing its core components. Initially, a green (532 nm) laser beam is shaped into a sheet and coupled through polished side facets of the diamond to illuminate and initialize NVs within the target FOV. Unlike conventional light-sheet systems~\cite{Olarte_LSM_2018}, the thickness of the light-sheet in the LC-QDM is not critical, provided that the entire region of interest is uniformly illuminated. After a sufficient initialization time $t_{init}^{LS}$ (typically 1-100 $\mu$s), the light-sheet is turned off. 

Next, microwave (MW) pulses are applied to the same NV population to perform quantum sensing protocols, delivered via a coaxial loop or resonator \cite{Eisenach2018loopgapresonator} to ensure field homogeneity across the FOV and compatibility with light-sheet illumination. 

Finally, a second green laser beam (can be from the same source as the first beam), tightly focused by a microscope objective, is utilized to read out individual NV sensing voxels ($\sim$1 $\mu$m$^3$). A pair of AOMs is used to rapidly switch the position of the readout laser beam on the target FOV and generate a QDM image, similar to a laser scanning confocal system (Fig. \ref{fig:1}b). At each position, the readout laser dwells for a readout time $t_{RO}^{conf}$ to capture sufficient NV PL signal (typically 1-10 $\mu$s). The values of the initialization and readout times ($t_{init}^{LS}$, $t_{RO}^{conf}$) depend on the laser intensity (see Section~\ref{sec:nvinitreadout}). The resulting NV PL is collected by the same readout objective and propagates along unique paths for different NV voxels. A second pair of AOMs is introduced to descan the PL to maintain alignment with a fixed pinhole, thus allowing confocal readout. The drive radiofrequencies (RFs) of the two AOM pairs are calibrated to accurately point the path of the readout laser beam and maximize PL transmission through the pinhole. (If needed, a double-pass configuration~\cite{donleyDoublepassAcoustoopticModulator2005} can be implemented for the readout AOMs to mitigate chromatic aberration effects arising from the broad NV PL spectrum.) Synchronizing the RF signals for both AOM pairs minimizes dead time during beam movement and enables signal averaging for enhanced sensitivity.

\section{Performance of the LC-QDM}
\label{sec:anticipatedperformance}

\subsection{Advantages of the LC-QDM}
The LC-QDM enables high-speed, diffraction-limited confocal readout of NV PL from a bulk diamond sample, integrating key advantages of both scanning confocal and 2D widefield QDMs. A potential alternative system that integrates speed and optical sectioning is a light-sheet QDM without confocal readout~\cite{Olarte_LSM_2018,Ang2024,Horsley_lightsheet_2018}. However, achieving $\sim$1 $\mu$m Z-resolution while maintaining a large FOV remains challenging with this system, with the best result to date achieving a Z-resolution of 14 $\mu$m~\cite{Horsley_lightsheet_2018}. Significant improvements with this approach will likely require the use of non-Gaussian beams for the light-sheet~\cite{Olarte_LSM_2018}, significantly increasing optical complexity. Additionally, the light-sheet profile tends to be distorted when positioned close to the diamond surface due to diamond edge clipping. In contrast, the LC-QDM utilizes a confocal pinhole to accomplish Z-sectioning with a resolution $\sim$1 $\mu$m (typical in confocal QDMs). As the sectioning is performed by the pinhole, a relatively simple, crude light-sheet suffices for global optical NV spin initialization.

In addition to optical Z-sectioning capability, the LC-QDM measurement protocol is more efficient than existing QDMs protocols. Using a light-sheet to initialize NVs in the entire FOV simultaneously allows a larger number of recurrent readouts within the NV $T_{1}$ limit compared to the method in Ref.~\cite{leibold2024time}. In particular, our protocol excels in applications with longer NV initialization and readout durations. These durations depend inversely on laser intensity, which can lead to practical constraints in real-world applications. For example, the intensity of a broad light-sheet laser beam can be limited by practically available laser power. Additionally, when a sample is placed directly on the top surface of the diamond chip, a common strategy to minimize sensor-to-sample standoff distance, the sample may be exposed to irradiation from the vertical readout beam. In bioimaging applications, such irradiation may cause phototoxicity in live cells~\cite{laissueAssessingPhototoxicityLive2017,xuSuperResolutionEnabledWidefield2024}. Such constraints limit the laser power, potentially lengthening the initialization and readout times to several tens of microseconds, degrading the sensitivity of traditional QDMs and lengthening data collection. In these applications, LC-QDM readout is significantly more efficient than other QDMs. Furthermore, with the LC-QDM, the laser light enters the diamond from the side and does not interact directly with the sample, providing another advantage in bioimaging applications compared to confocal or total internal reflection illumination in widefield QDMs.

\subsection{LC-QDM readout efficiency}
\label{sec:nvinitreadout}

\begin{figure}
    \centering
    \includegraphics[width=6in,
  keepaspectratio,]{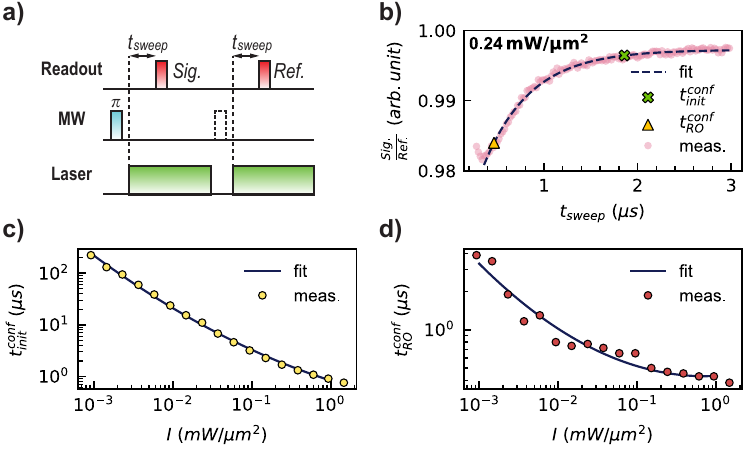}
    \caption{Measured dependence of NV initialization and readout durations on laser intensity using a confocal QDM. a) Measurement pulse sequence. Following a few $\mu$s laser pulse to initialize the NV spins to $m_s$ = 0 (not shown) and a MW $\pi$ pulse to invert the NVs to $m_s$ = +1, the readout data acquisition system records the instantaneous NV PL at a variable delay ($t_{sweep}$) after the laser is turned back on for another few $\mu$s. This signal PL measurement (Sig.) is followed by a reference PL measurement (Ref.) for the same delay but without the MW $\pi$ pulse. Sweeping this delay enables simultaneous determination of optimal laser duration for NV spin-state readout and polarization. b) Example measurement data employing pulse sequence in a). See main text for discussion of the determination of NV spin readout $t_{RO}^{conf}$ and initialization $t_{init}^{conf}$ times. c) Experimentally determined $t_{init}^{conf}$ as a function of laser intensity $I$. Laser intensity is extracted from the measured power incident on the diamond divided by the confocal collection area. We assume the laser has a symmetric Gaussian profile. Uniform illuminance of the NVs is ensured by utilizing a confocal QDM with a pinhole whose size when projected to the objective plane is 2.5\,$\times$ smaller than beam waist of the focused green laser. Data is fitted to a second-order polynomial function on a logarithmic scale. d) Experimentally determined $t_{RO}^{conf}$ as a function of laser intensity $I$, determined as in c). Measurements presented here use a CVD-grown diamond chip with the following properties: 3 mm x 3 mm x 0.5 mm; 99.99\% $^{12}$C; bulk NV diamond with nominal [NV] = 0.3\,ppm. 
    } 
    \label{fig:2}
\end{figure}

To quantify the LC-QDM's projected readout efficiency advantage in more detail, we first measure the dependence of optical initialization and readout durations on laser intensity (Fig. \ref{fig:2}). The NV is an electronic spin-1 system that emits PL brighter in the $m_{s}=0$ state compared to $m_{s}=\pm1$ when undergoing optical pumping. We measure the NV PL contrast between $m_s$ = 0 and $m_s$ = +1 on a confocal QDM (no light-sheet) using the measurement protocol shown in Fig. \ref{fig:2}a. From the measured dependence of contrast on the laser illumination duration $t_{\mathrm{sweep}}$ (Fig. \ref{fig:2}b), we determine i) $t_{RO}^{conf}$ from the $t_{\mathrm{sweep}}$ value that maximizes the product of contrast and the square root of PL photon flux; and ii) $t_{init}^{conf}$ from the $t_{\mathrm{sweep}}$ value for which contrast decays to $1/e^3$ of its peak value, effectively achieving $\gtrsim$\,95\,\% initialization efficiency with an infinitely long laser pulse. Subsequently, we experimentally determine $t_{init}^{conf}$ and $t_{RO}^{conf}$ at different laser intensities $I$ (Figs.~\ref{fig:2}c, d). Fitting the data to exponentials allows us to extract the dependence of $t_{RO}^{conf}$ and $t_{RO}^{conf}$ on $I$.

\begin{figure}
    \centering
    \includegraphics[width=\linewidth]{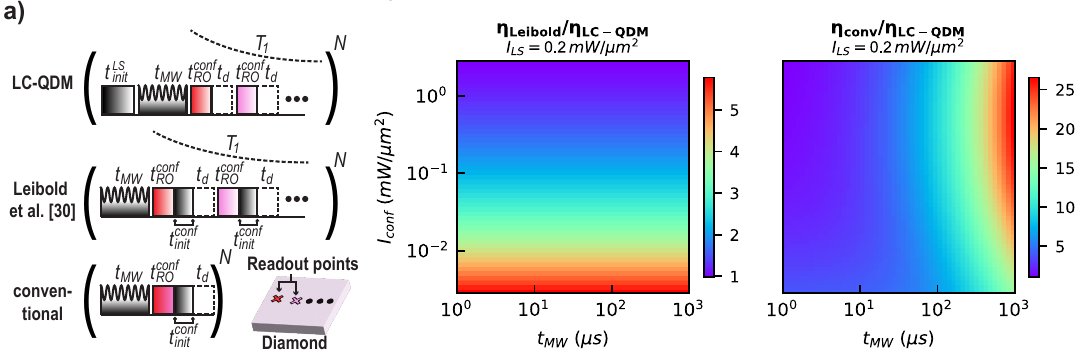}
    \caption{Sensitivity comparison between LC-QDM, scanning QDM of Ref.~\cite{leibold2024time}, and a conventional scanning QDM. a) Schematic of measurement protocols employed in different QDMs. The LC-QDM employs global NV spin initialization, MW control, and recurrent local readout from as many spatial points as possible before $T_{1}$ limit. In comparison, Ref.~\cite{leibold2024time} utilizes global MW control with recurrent local readout and spin initialization; and a conventional scanning QDM utilizes local MW control, initialization, and readout. See main text for details. b) Calculation of relative per-voxel measurement sensitivity $\eta$ as a function of readout laser intensity and MW pulse sequence duration, assuming $t^{LS}_{init} \approx t^{conf}_{init}$. Note that smaller $\eta$ indicates a more sensitive measurement. Plots shown are for a fixed light-sheet intensity $I_{LS}=0.2$\,mW/$\mu$m$^2$. LC-QDM is projected to outperform the scanning QDM of Ref.~\cite{leibold2024time} and the conventional scanning QDM for nearly all experimental parameters simulated here.
    } 
    \label{fig:3}
\end{figure}

Next, we apply these results to compare the projected readout efficiency of the LC-QDM to other scanning QDMs (Fig. \ref{fig:3}a). Readout efficiency can be quantified using the per-voxel sensitivity $\eta$ to the parameter of interest (e.g., magnetic field), which relates signal-to-noise ratio (SNR) and total time $t$ required to initialize and read out a single sensing voxel (including optical initialization and MW control sequence duration) as follows:
\begin{equation}
\eta=\frac{1}{SNR}\times\sqrt{t}.
\end{equation}
The LC-QDM employs recurrent readout to amortize the time cost of initialization and MW control sequence. The number of maximum readouts is limited by NV $T_{1}$ due to thermalization erasing the sensed information encoded in the NV quantum state distribution. If we assume a normalized readout SNR of 1 at the beginning of $T_{1}$ decay, a global initialization (via light-sheet) time of $t_{init}^{LS}$, a MW sequence time of $t_{MW}$, a single-voxel confocal readout time of $t_{RO}^{conf}$ and a beam movement dead-time of $t_{d}$, then $\eta$ can be approximated as: 
\begin{equation}
       \eta_{LC-QDM}=\underbrace{\frac{2}{1+e^{-1}}}_{\substack{\text{Avg. SNR between first} \\ \text{and last readout at $T_{1}$}}}
\times\underbrace{\sqrt{\frac{(t_{init}^{LS}+t_{MW}+T_{1})(t_{RO}^{conf}+t_{d})}{T_{1}}}}_{\substack{\text{Avg. time required for} \\ \text{each recurrent readout}}}.
    \label{eq:eta_lcqdm}
\end{equation}
In comparison, the scanning QDM of Ref.~\cite{leibold2024time} does not have global NV initialization. Instead, after a global MW control sequence, a laser is applied to read out the NVs (lasting for $t_{RO}^{conf}$) and then to reinitialize them (lasting for $t_{init}^{conf}$) at each sensing voxel in a recurrent fashion until $T_{1}$ (Fig.~\ref{fig:3}a). The per-voxel sensitivity from this protocol can be expressed as:
\begin{equation}
    \eta_{Leibold}=\frac{2}{1+e^{-1}}\times\sqrt{\frac{(t_{MW}+T_{1})(t_{RO}^{conf}+t_{init}^{conf}+t_{d})}{T_{1}}}.
    \label{eq:eta_leibold}
\end{equation}
For a conventional scanning QDM, only a single voxel is read out after initialization and a MW sequence. SNR is maximized for each NV sensing voxel by trading off time required for readout, resulting in per-voxel sensitivity given by:
\begin{equation}
\eta_{conv}=\sqrt{t_{MW}+t_{RO}^{conf}+t_{init}^{conf}+t_{d}}.
\end{equation}

\subsection{LC-QDM performance comparison}

We use the above measurements and expressions to estimate the relative per-voxel sensitivity of the LC-QDM to other scanning QDMs over a range of experimentally realistic values of $t_{MW}$, $I_{LS}$, and $I_{conf}$ (Fig. \ref{fig:3}b). The maximum light-sheet intensity is assumed to be $2$\,mW/$\mu$m$^2$, which can be achieved with 2\,W laser power and a light-sheet thickness of 10\,$\mu$m and a lateral size of 100\,$\mu$m. The minimum readout laser power is $2$ $\mu$W, which could be required in imaging applications with live cells to reduce phototoxicity. The MW sequence duration depends on diamond material properties \cite{Bauch2020} and measurement protocols \cite{levine_principles_2019}; a range between 1\,$\mu$s to 1\,ms is chosen for the present performance comparison. A complete list of parameter choices is given in Table \ref{table:1}.
For the parameter space used here, the LC-QDM provides the best per-voxel sensitivity due to the long NV $T_{1}$ and increased readout duty cycle. The LC-QDM outperforms a conventional scanning QDM by an order-of-magnitude due to the lack of recurrent readout for the latter. The improvement is modest (compared to Ref.~\cite{leibold2024time}) when the confocal readout laser intensity is close to the NV saturation intensity ($\sim$1\,mW/$\mu$m$^2$, where $t_{RO}^{conf}\sim t_{init}^{conf}$). However, in the case of low laser intensity ($\sim$100 $\mu$W/$\mu$m$^2$, $t_{RO}^{conf}\ll t_{init}^{conf}$), the LC-QDM becomes increasingly advantageous; e.g. a 5\,$\times$ improvement of per-voxel sensitivity corresponds to 25\,$\times$ reduction of experimental time. We note that the behavior shown in Figure \ref{fig:3}b remains nearly unchanged for all light-sheet intensities used in the modeling as $T_{1}$ dominates over $t_{init}^{LS}$ even for the weakest $I_{LS}$. Thus, the enhanced per-voxel sensitivity of the LC-QDM (Eq. \ref{eq:eta_lcqdm}) relative to the scanning QDM of Ref.~\cite{leibold2024time} (Eq. \ref{eq:eta_leibold}) arises primarily from differences between $t_{RO}^{conf}$ and $t_{init}^{conf}$, as the readout duration has less-dependence on laser intensity (see Figs. \ref{fig:2}c, d).

\begin{table}
\begin{center}
\begin{tabular}{ c c c }
% \centering
 Symbol & Description & Value\\
 \hline
 \hline
 $L_{y}$ & Light-sheet dimension in $y$ direction & 100\,$\mu$m \\  
 $d_{LS}$ & Light-sheet thickness & 10\,$\mu$m \\
 $P_{LS}$ & Laser power for light-sheet & 2$-$2000\,mW \\
 $I_{LS}$ & Intensity of light-sheet within Rayleigh range & $I_{LS}=\frac{P_{LS}}{L_{y}\times d_{LS}}$ \\
 $\delta_{conf}$ & Confocal readout laser beam diameter at focus & 0.53\,$\mu$m \\
 $P_{conf}$ & Laser power for confocal readout & 2\,$\mu$W(bio.)$-$2\,mW \\
 
 $I_{conf}$ & Readout laser intensity within Rayleigh range & $I_{LS}=\frac{P_{LS}}{\delta_{conf}^2}$\\
 $t_{init}^{LS}$ & Spin initialization time at focus of light-sheet & Fig. \ref{fig:2}c \\
 $t_{init}^{conf}$ & Spin initialization time at focus of readout laser &  Fig. \ref{fig:2}c\\
 $t_{RO}^{conf}$ & Spin readout time at focus of readout laser &  Fig. \ref{fig:2}d\\
 $t_{d}$ & Dead time for steering readout laser and PL & 100\,ns \\
 $t_{MW}$ & MW sequence duration & 1$-$1000\,$\mu$s \\
 
\end{tabular}
\end{center}
\caption{Parameters used for comparison of relative per-voxel sensitivity shown in Figure \ref{fig:3}.}
\label{table:1}
\end{table}

\section{Conclusion}
We outline the design, operation, and projected performance of the light-sheet confocal quantum diamond microscope (LC-QDM) to enable rapid readout of large NV sensor volumes while maintaining diffraction-limited resolution in 3D. The LC-QDM has two key features. First, a broad light-sheet is used for initialization of all NVs across a wide FOV. Second, a pair of AOMs steers the readout laser beam so that a fixed pinhole can reject out-of-focus light to achieve confocal resolution of NV PL. This inherent high-resolution readout obviates the need for diamond chips with thin NV layers or complex illumination beam shaping. Instead, a commercially available bulk NV-diamond chip can be used for efficient, widefield 3D quantum sensing with spatial resolution $\sim$1 $\mu$m. Decoupling NV spin initialization from readout is particularly advantageous for applications where readout laser power is limited, such as magnetic imaging of biological samples. The LC-QDM should be compatible with diverse NV measurement protocols \cite{Tang2023QDM,marshall_high-precision_2022,glennHighresolutionMagneticResonance2018}, making it useful for a host of quantum sensing applications in areas such as material science, strain imaging, damage track detection, and bioimaging.

\section*{Acknowledgments}

We thank Johannes Cremer, Connor Hart, Stephen DeVience, Mason Camp, Maximilian Shen, Jiarui Yu, Richard Escalante, and Andrew Beling for useful discussions. This work is supported by, or in part by, the U.S. Army Research Laboratory under Contract No. W911NF2420143; the U.S. Army Research Office under Grant No. W911NF2120110; and the University of Maryland Quantum Technology Center.

\section*{Author Contributions}
Daniel G. Ang and Jiashen Tang contributed equally to this work. Conceptualization, methodology, investigation and validation: D.G. Ang and J. Tang; editing: D.G. Ang, J. Tang, and R.L. Walsworth; supervision and project administration: D.G. Ang and R.L. Walsworth; funding acquisition: R.L. Walsworth. All authors have read and agreed to the published version of the manuscript.

% \section*{Funding Statement}
% This research received no external funding.

\section*{Conflict of Interest}

R.L. Walsworth is a founder of and advisor to companies that are developing and commercializing NV sensing technology. These relationships are disclosed to and managed by the University of Maryland Conflict of Interest Office.

% \section*{Competing Interest}
% Not applicable.

% \section*{Institutional Review Board Statement}
% Not applicable.

\section*{Data Availability}

The data that support the findings of this study are available from the corresponding authors upon reasonable request.

% ----------------------------
% References
% ----------------------------
\bibliographystyle{JHEP}  % or unsrt, apsrev4-2, etc.
\bibliography{reference}

\end{document}